\documentclass[aps,prl,twocolumn,showpacs]{revtex4-1}
\usepackage{graphicx}
\usepackage{bm}
\usepackage{color}

\begin{document}

\title{Burnett Coefficients in Quantum Many-Body Systems}
\author{R. Steinigeweg$^{1,2}$}
\affiliation{$^1$Institute for Theoretical Physics, Technical University Braunschweig, D-38106 Braunschweig, Germany\\ $^2$J. Stefan Institute, SI-1000 Ljubljana, Slovenia}
\author{T. Prosen$^3$}
\affiliation{$^3$Faculty of Mathematics and Physics, University of Ljubljana, SI-1000 Ljubljana, Slovenia}

\date{\today}

\begin{abstract}
The Burnett coefficient $B$ is investigated for transport in one-dimensional
quantum many-body systems. Extensive numerical computations in spin-$1/2$ chains
suggest a linear growth with time, $B(t) \sim t$, for non-integrable chains
exhibiting diffusive transport. For integrable spin chains in the metallic
regime, on the other hand, we find a cubic growth with time, $B(t) \sim
-D_{\rm m}^2 t^3$, with the proportionality constant being simply a square of
the Drude weight $D_{\rm m}$. The results are corroborated with additional
studies in non-interacting quantum chains and in the classical limit of
large-spin chains.
\end{abstract}


\maketitle

{\it Introduction.}---
Understanding classical and quantum diffusion in deterministic Hamiltonian
systems is one of the most ubiquitous problems of statistical physics
\cite{gaspard1998}. In Fourier space of momentum $q$, diffusion is described by
the well-known equation
\begin{equation}
\dot{\rho}_q(t
) = -q^2 D(t) \, \rho_q(t) \, , \quad D(t) = D
\end{equation}
and manifests as the simple exponential relaxation of a harmonic density
profile at a characteristic time scale $\tau = 1/q^2 D$. The strict derivation
of diffusion from truly microscopic principles remains a challenge to theorists \cite{bonetto2000},
and the problem is often simplified to a mere calculation of the diffusion coefficient
$D(t)$ in the limit $q=0$ via the famous Green-Kubo formula \cite{kubo1991}. It has become clear
that $D(t)$ can diverge in integrable systems \cite{prosen2011}, $D(t) \propto t$ \cite{steinigeweg2011-1}, due to the lack
of sufficient scattering, which is a key issue for understanding large thermal spin
transport in quantum magnets \cite{sologubenko2000} or thermalization in cold atomic gases \cite{cazalilla2010}. On the
other hand, $D(t)$ is believed to be constant, $D(t) = D$, in generic nonintegrable
systems as a consequence of microscopic Hamiltonian chaos \cite{gaspard1998}. This believe opens the
important question whether diffusion is rather the rule than the exception.
\\
The existence of diffusion can only be clarified by taking into account finite
momentum $q \neq 0$ \cite{sirker2009,steinigeweg2011-2}. The first higher order correction can be systematically
described by the so-called Burnett coefficient $B(t)$ \cite{gaspard1998,ernst1981,bijeren1982},
\begin{equation}
\dot{\rho}_q(t) = [-q^2 D(t) + q^4 B(t) + \ldots ] \, \rho_q(t) \, , \label{expansion}
\end{equation}
which may diverge even for dynamical processes with a constant $D(t)$ \cite{ernst1981}.
Even though Burnett coefficients have been extensively studied in the literature for
various classical models, in particular for Lorentz-type gases \cite{bijeren1982,chernov2000},
essentially nothing is known about Burnett coefficients in quantum systems.
\\
In this Letter we do the first steps by calculating $B(t)$ numerically  for various
one-dimensional, integrable and nonintegrable models, including spin-$1/2$ $XXZ$
chains, large-spin chains, and more abstract models of quantum transport. We
generally observe the moderate divergence $B(t) \sim B' \, t$, for
cases with a constant $D(t) \sim D$. At the characteristic time scale $\tau$, this
observation implies
\begin{equation}
\dot{\rho}_q(t = \tau) = [-q^2 D + q^2 B'/D + \ldots ] \, \rho_q(t = \tau) \, ,
\end{equation}
i.e., Burnett coefficients are a relevant correction at any finite momentum $q \neq
0$, which speed up or slow down the still diffusive relaxation (if $|B'| \lesssim D^2$).
For ballistic cases with $D(t) \sim D_{\rm m} \, t$, on the other hand, we find the
stronger divergence $B(t) \sim -D_{\rm m}^2 t^3$ with the Drude weight $D_{\rm m}$ 
as the constant of proportionality.

\begin{figure}[tb]
\includegraphics[width=0.80\columnwidth]{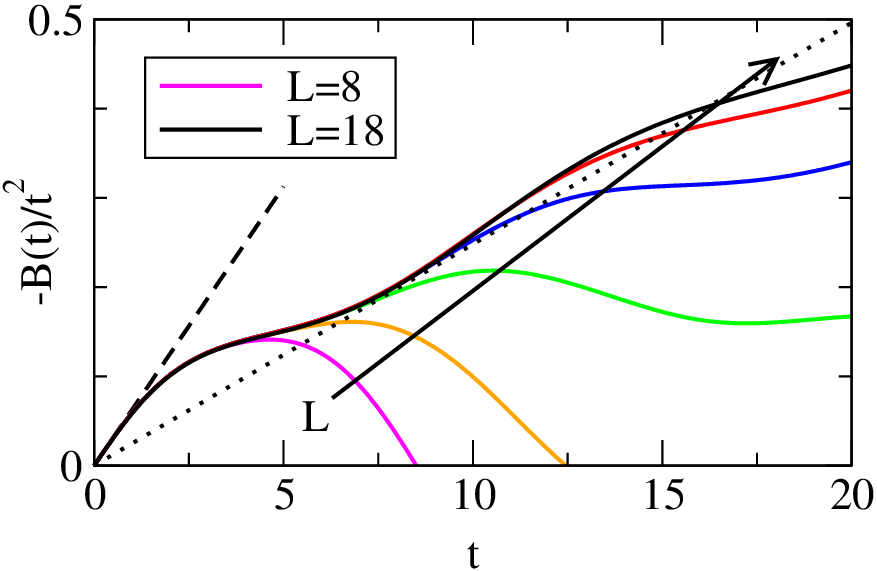}
\caption{(color online) Burnett coefficient $B(t)$, divided by $t^2$, for the
spin-1/2 $XXZ$ chain at $\Delta=0.5$ for $L=8,10,\ldots,18$ and all magnetization
sectors. At short times or at $\Delta = 0$ (free fermions), $B(t)/t^2$ is given
by the function $-t/16$ (dashed curve). In addition, a function
$-(0.63)^2 t/16$ is indicated (dotted curve), see prediction (\ref{Drude}).}
\label{FigI}
\end{figure}

{\it Diffusion and Burnett coefficient.}---
Following \cite{ernst1981,lowe1997,chernov2000} the quantum
Burnett coefficient $B(t)$ may be introduced by formally expanding the decay rate
of density-density correlation functions in $q$. This expansion leads to
the time-dependent diffusion coefficient
\begin{equation}
D(t) = \frac{1}{\chi} \int \limits_0^t \! \mathrm{d}t_1 \,\, f(t_1) \, , \quad
f(t_1) = \langle J(0) J(t_1) \rangle \, , \label{D}
\end{equation}
given as a time integral over the two-point correlation function of the
current operator $J(t)$ in the Heisenberg picture, where $\langle \bullet
\rangle = {\rm tr}(\bullet)/{\rm dim}$ denotes an equilibrium expectation at high
temperatures, as considered in this Letter. The further occurring
prefactor $\chi$ is a constant and denotes the ``static susceptibility''
\cite{susceptibility}.
\\
The time-dependent Burnett coefficient is the {\it
difference} between two contributions,
\begin{equation}
B(t) = I_4(t) - I_{2-2}(t) \, , \label{B}
\end{equation}
where the first term $I_4(t)$ is given by
\begin{equation}
I_4(t) = \frac{2}{\chi} \int \limits_{0}^t \! \mathrm{d}t_1 \int \limits_{0}^{t_1}
\! \mathrm{d}t_2 \int \limits_{0}^{t_2} \! \mathrm{d}t_3 \,\, \langle J(0) J(t_1)
J(t_2) J(t_3) \rangle
\end{equation}
as a triple-time integral over the {\it time-ordered} four-point current
autocorrelation function. The second term
\begin{eqnarray}
I_{2-2}(t) &=& \frac{2}{\chi} \int \limits_{0}^t \! \text{d}t_1 \int
\limits_{0}^{t_1} \! \text{d}t_2 \int \limits_{0}^{t_2} \! \text{d}t_3 \,\,
[f(t_1) f(t_2-t_3) \nonumber\\
&+& f(t_2) f(t_1-t_3) + f(t_3) f(t_1-t_2)]
\end{eqnarray}
is a similar time-ordered integral but over products of two-point correlations
\cite{algorithm}. Conveniently, this contribution can be rewritten as
$I_{2-2}(t) = 2 \chi D(t) \int_0^ t \mathrm{d}t_1 D(t_1)$, particularly
unveiling the linear increase $I_{2-2}(t) \propto t$ in the case of a existent
diffusion constant. But, despite the apparent divergence of $I_{2-2}(t)$ in
that case, the Burnett coefficient can still remain finite, as discussed in the
following. To this end, assume for the moment that (i) the two-point
correlation $f(t)$ is a delta function $\delta(t)$ and that (ii) the four-point
correlation $\langle J(0) J(t_1) J(t_2) J(t_3) \rangle$ can be factorized as
$f(t_3) f(t_1-t_2)$. Then the contributions $I_{2-2}(t)$ and $I_4(t)$ are
identical and, as a consequence, the Burnett coefficient vanishes exactly.
Giving up the assumption (i) by broadening the delta function still allows for
a finite Burnett coefficient. While the assumption (ii) appears to be crucial,
it may be fulfilled for a non-integrable model with $J$, in the energy
eigenbasis, being a random Hermitian matrix and, consequently, $J^2$ being
close to proportional to an identity matrix.

\begin{figure}[tb]
\includegraphics[width=0.85\columnwidth]{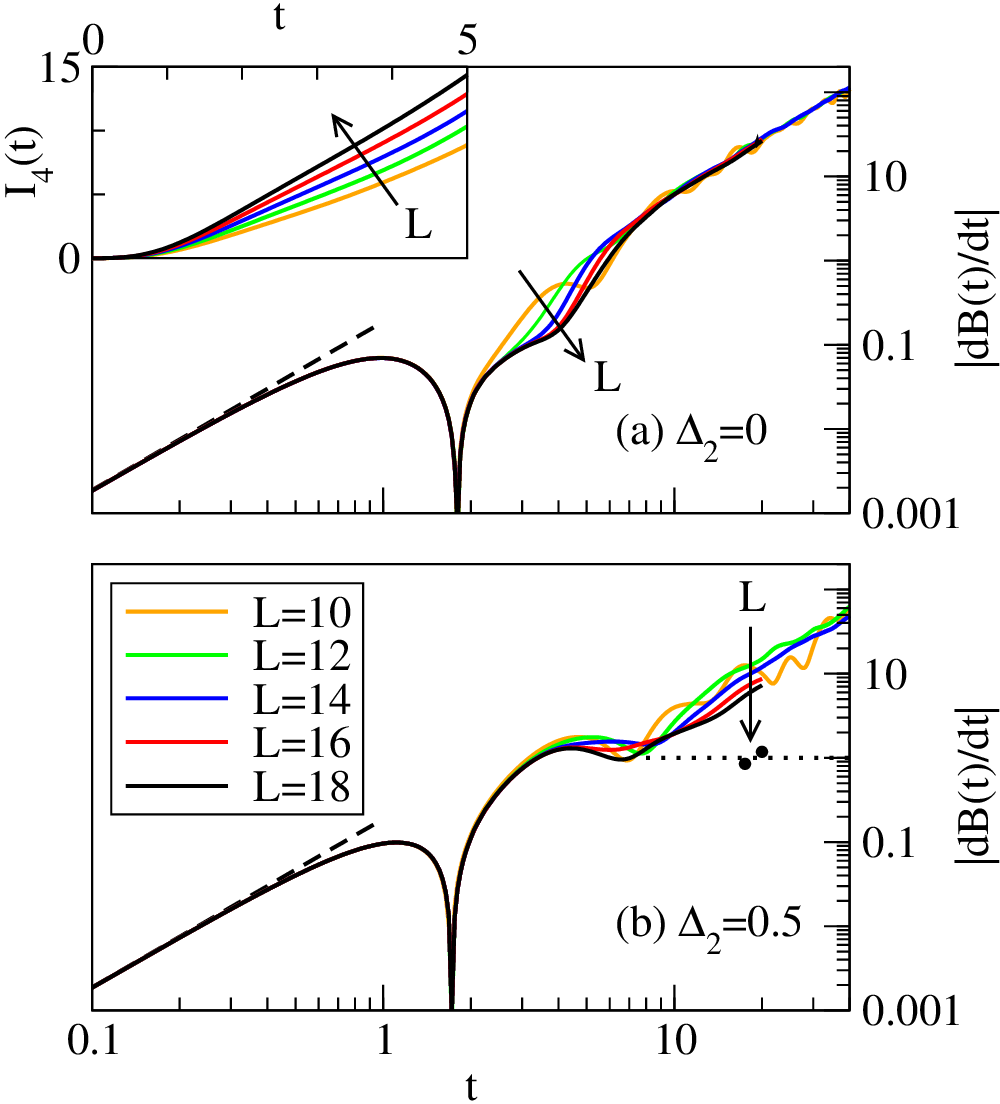}
\caption{(color online) Absolute value of the derivated Burnett coefficient,
$|\mathrm{d}B(t)/\mathrm{d}t|$, for the spin-$1/2$ $XXZ$ chain at $\Delta=1.5$
(a), and for its non-integrable modification with $\Delta_2=0.5$ (b). Numerical
results (solid curves) for $L=10, 12, \ldots, 18$ are shown in a log-log plot.
In addition to the short-time behavior (dashed curves), the long-time behavior
is extrapolated in (b) using the observed exponential scaling with $L$ (dotted
curve). Inset: The contribution $I_4(t)$ by itself increases with $L$.}
\label{FigII}
\end{figure}

{\it Anisotropic spin-$1/2$ Heisenberg chain.}---
We are going to investigate the transport of magnetization in the spin-$1/2$
$XXZ$ model as a paradigmatic example of an interacting {\it many}-particle
quantum system in one dimension. The $XXZ$ Hamiltonian is given by
\begin{equation}
H = \sum_{r=1}^L (S_r^x S_{r+1}^x + S_r^y S_{r+1}^y + \Delta \, S_r^z
S_{r+1}^z) \, ,
\end{equation}
where $S_r^{x,y,z}$ are the components of spin-$1/2$ operators at site $r$, $L$
is the number of sites arranged periodically, $L+1\equiv 1$, and $\Delta$ is the
anisotropy. The magnetization current
\begin{equation}
J = \sum_{r=1}^L (S_r^x S_{r+1}^y - S_r^y S_{r+1}^x)
\end{equation}
commutes with $H$ in the non-interacting case $\Delta=0$. In that case (due to
$\langle J^2 \rangle = L/8$, $\langle J^4 \rangle = 3(L^2-L)/64$, and $\chi =
L/4$), one obtains directly $D(t) = t/2$ and $B(t)=-t^3/16$, which for $\Delta
> 0$ remains only an approximation at short times, in agreement with Eq.~(4) of
Ref.~\onlinecite{steinigeweg2011-2}. Remarkably, at $\Delta = 0$ a series
expansion of density autocorrelations (Bessel functions \cite{fabricius1997})
leads also to $q^2 D(t)$ and $-q^4 B(t)$ as the leading terms, hence being a
convincing consistency check of the present approach.

In the metallic regime, $0 < |\Delta| < 1$, the magnetization current is still
partially conserved such that the two-point correlation $f(t)$ does not decay
to zero but remains at a finite Drude weight $D_\mathrm{m}=\lim_{t\to\infty}
f(t)$, $0 < D_\mathrm{m} < 1/8$, recently proven by establishing positive lower
bounds in the thermodynamic limit \cite{prosen2011}. This finite Drude weight
implies the linear increase of the diffusion coefficient at long times, $D(t)
\propto 4 D_\mathrm{m} t$, just as in the case of $\Delta = 0$. Factoring the
four-point correlation at long times \cite{prosen02}, one derives the
asymptotics of the Burnett coefficient as
\begin{equation}
B(t) \simeq -4 D_\mathrm{m}^2 t^3 \, .
\label{Drude}
\end{equation}
In Fig.~\ref{FigI} we demonstrate this result by numerically simulating $B(t)$
for finite length $L =10, \ldots, 18$ using {\it all} invariant subspaces
(translation, magnetization) and also using 4th order Runge-Kutta integration
for generating time order \cite{algorithm} (step size $\delta t =0.02$).
While Fig.~\ref{FigI}
clearly shows for $\Delta = 0.5$ a stronger than quadratic increase of $B(t)$
with time, it also is consistent with $B(t) \sim -(0.63)^2 t^3/16$ at long
times, e.g., $63 \%$ of the Drude weight in the case of $\Delta = 0$.
Remarkably, the exact Drude weight at $\Delta = 0.5$ for finite $L=18$ $[20]$
is $63 \%$ $[62 \% ]$ of the Drude weight at $\Delta = 0$, while the lower
bound in the thermodynamic limit is $56 \%$ \cite{prosen2011}.

Eventually, we discuss the regime $\Delta > 1$, where Drude weights are
expected to vanish in the thermodynamic limit and strong evidence of
magnetization diffusion has been found in non-equilibrium bath scenarios on the
basis of the Lindblad equation \cite{michel2008,znidaric2011}. The diffusion
coefficient has been shown to behave as $D(t > 3.0/\Delta) \approx 0.88/
\Delta$ at vanishing \cite{steinigeweg2011-1} and finite momentum
\cite{steinigeweg2011-2}. We focus on the Burnett coefficient $B(t)$ in this
Letter. Figure \ref{FigII} (a) shows numerical results summarizing an order of a
CPU-year of simulations and plotting, for convenience, $|\mathrm{d}B(t)/
\mathrm{d}t|$ in a log-log scale. Several comments are in order: First, after
the already discussed $t^3$ behavior at short times, the Burnett coefficient
changes its sign, visible as the zero drop in Fig.~\ref{FigII} (a), and
indicates a correction towards an insulator. Second, curves for $L \geq 12$
have converged for times after the zero drop and show at least the tendency to
form a plateau at $t \sim 4$ for $L \rightarrow \infty$, then indicating a
linearly increasing Burnett coefficient $B(t) \propto t$. And third, even
though a possible plateau is not pronounced yet, the contributions $I_4(t)$ and
$I_{2-2}(t)$ increase linearly with time at $t \sim 4$, see the inset of
Fig.~\ref{FigII} (a). Notably, the Burnett coefficient turns out to be the {\it
small} difference between both contributions, which by themselves diverge with
$L$. This divergence does not appear in the modular quantum system,
discussed later. Finally, we show in Fig.~\ref{FigII} (b) numerical results
for the modified {\it non-integrable} $XXZ$ model $H + \Delta_2 \sum_r
S^z_r S^z_{r+2}$ with $\Delta_2 = 0.5$. While the overall structure of
$|\mathrm{d}B(t)/\mathrm{d}t|$ is similar, an emerging plateau is more clearly
visible for the accessible lengths, hence pointing towards a linearly
increasing Burnett coefficient again. Summarizing, a linear asymptotic scaling
$B(t)\propto t$ is clearly suggested in either integrable or non-integrable
regimes with a finite diffusion constant.

\begin{figure}[tb]
\includegraphics[width=0.80\columnwidth]{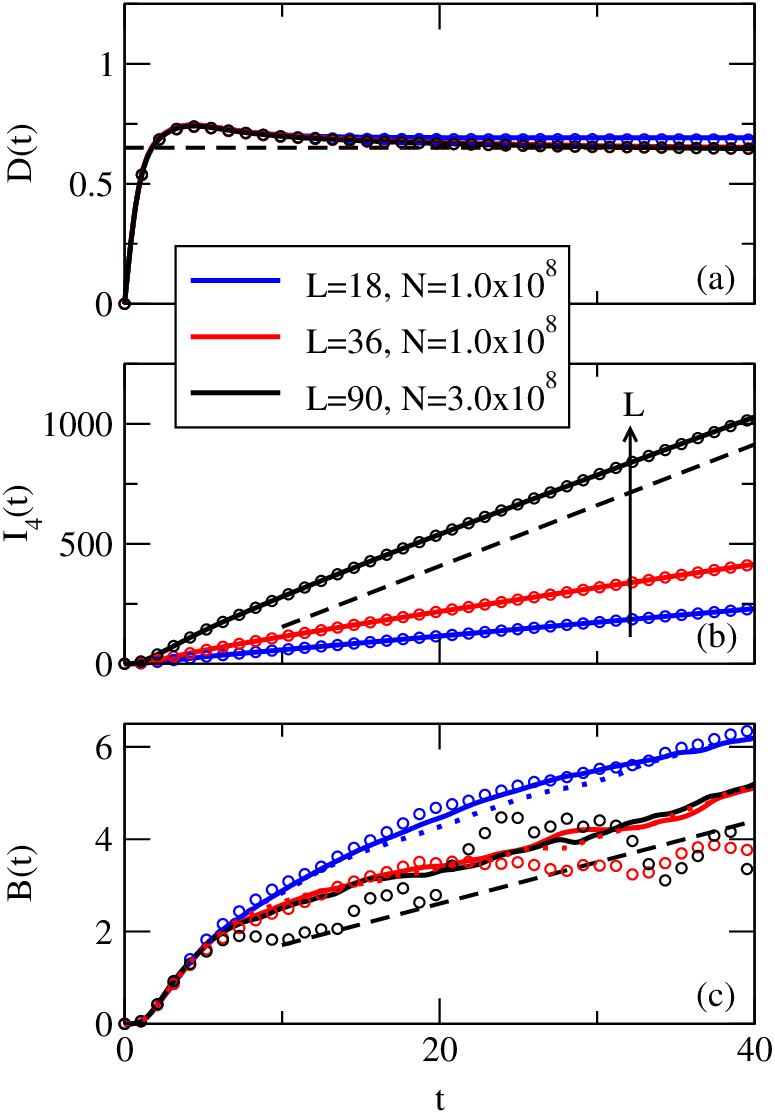}
\caption{(color online) Classical (a) diffusion coefficient $D(t)$, (b) the
contribution $I_4(t)$, and (c) Burnett coefficient $B(t)$ for the $XXZ$ model
at $\Delta = 1.5$ with $\Delta_2 = 0.5$. Numerical results (solid curves) on
finite length $L=18$, $36$, $90$ are obtained by numerically integrating
the Hamiltonian equations of motion (fixed step size $\delta t=0.05$) and by
averaging over $N \sim 10^8$ completely random initial states. The average
over only $N=10^7 \ll 10^8$ is also indicated (symbols). In (c) results
for a $2.5\times$ smaller time step are shown (dotted curves). Straight
lines (dashed curves) serve as guides to the eye.}
\label{FigIII}
\end{figure}

{\it Heisenberg chains in the large-spin limit.}---
In addition we present results on the classical limit of the considered spin
chains, where we focus on the case $\Delta = 1.5$ and $\Delta_2 = 0.5$ only. In
that limit the magnetization current is a function of classical unit (angular
momentum) vectors. Their dynamics we obtain by numerically integrating the
corresponding Hamiltonian equations of motion. Formally, the diffusion coefficient
in Eq.~(\ref{D}) and the Burnett coefficient in Eq.~(\ref{B}) remain defined the
same way, but the equilibrium average at high temperatures is now performed by
sampling over a set of completely random initial configurations, see
Ref.~\cite{mueller1988} for instance. While the required number of initial
states decreases with the chain length $L$ for the evaluation of the diffusion
coefficient \cite{steinigeweg2012}, the situation turns out to be different for
the evaluation of the Burnett coefficient. As in the quantum case, the Burnett
coefficient is the {\it small} difference between two contributions, which by
themselves diverge with $L$. Thus, errors due to insufficient averaging increase
with $L$ and can only be compensated by taking into account more and more
initial configurations. Approximately $N\approx10^8$ initial states are already
required for a chain of length $L=90$ taking about a CPU-year of computation
time. In Fig.~\ref{FigIII} we summarize these results for three different sizes
$L=18$, $36$, $90$. Apparently, $D(t \gtrsim 10)$ becomes independent of time for
all considered lengths $L \geq 18$. Remarkably, the quantitative value $D \approx
0.65$ is close to the expectation for the diffusion constant in the quantum
case \cite{steinigeweg2012}. Furthermore, $B(t \gtrsim 10)$ is observed to
increase linearly in time, at least for the largest two sizes $L \geq 36$. The
latter observation indicates the linear divergence of the Burnett coefficient
in diffusive classical spin chains. This is in clear agreement with the
finite-size results in the quantum case, see Fig.~\ref{FigII}, indicating that
the underlying mechanism ``survives'' the transition to the classical limit.

\begin{figure}[tb]
\includegraphics[width=1.0\columnwidth]{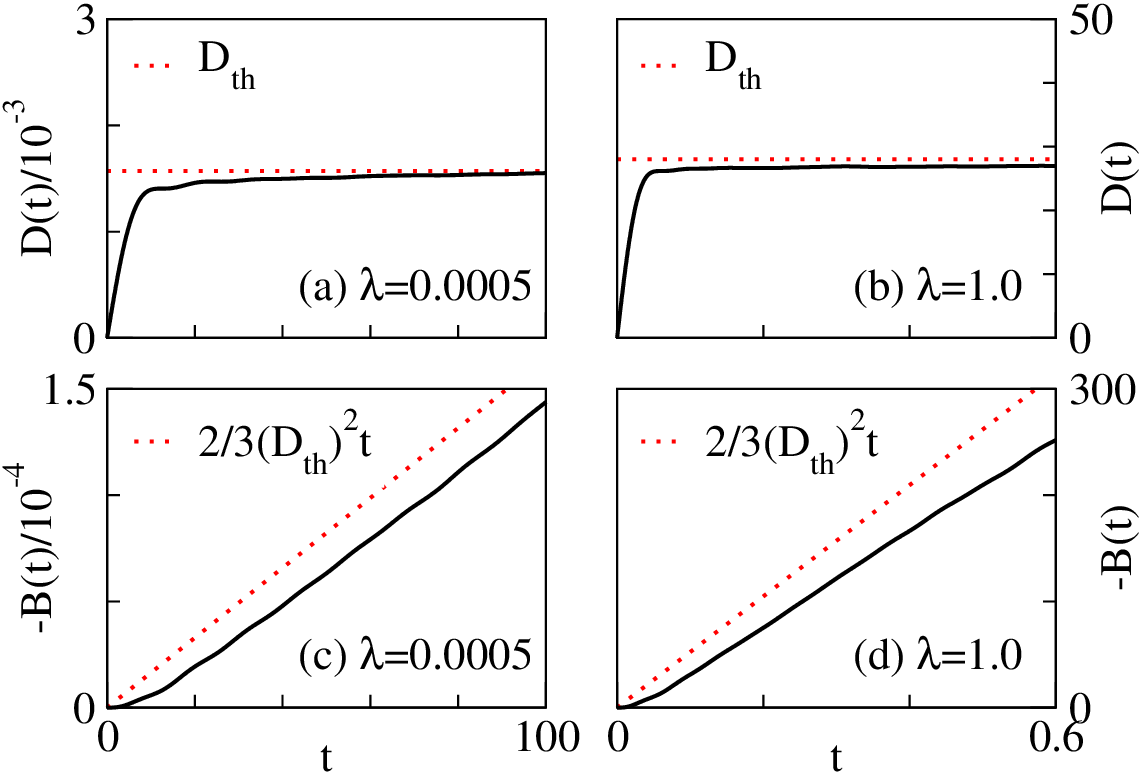}
\caption{(color online) Diffusion coefficient $D(t)$ and Burnett coefficient
$B(t)$ for the modular quantum system at coupling (a),(c) $\lambda = 0.0005$
and (b),(d) $\lambda = 1.0$. (Other parameters: $n=500$, $\delta \epsilon=0.5$.)
Numerical results (solid curves) on arbitrary length $L$ are shown for a {\it
representative} translation subspace (momentum $k=\pi/5$). Results agree well
with the theoretical predictions (non-solid curves).}
\label{FigIV}
\end{figure}

{\it ``Modular quantum system''.}---
Due to a delicate counterbalance of the terms $I_4$ and $I_{2-2}$, the
calculation of Burnett coefficients for many-body systems is extremely
demanding. Thus, in order to corroborate our prediction that $B(t)\propto t$
for one-dimensional lattice systems with finite diffusion constants, we make
another numerical experiment in a single-particle diffusive quantum system -- the
so-called {\em modular quantum system} \cite{steinigeweg2007,steinigeweg2011-1}.
Each of the $L$ local modules features an identical spectrum, consisting of $n$
equidistant levels in a band with the width $\delta \epsilon$. Therefore the
local Hamiltonian at the position $r$ is given by $h_r = \sum_\mu \mu \, \delta
\epsilon/n \, |r,\mu\rangle\langle r,\mu|$ in the one-particle basis $|r, \mu
\rangle$. The nearest-neighbor interaction between two local modules at the
positions $r$ and $r+1$ is $v_r = \lambda \, m_r + \text{H.c.}$,
\begin{equation}
m_r = \sum_{\mu,\nu} c_{\mu,\nu} \, |r, \mu \rangle \langle r+1, \nu |
\end{equation}
with the overall coupling strength $\lambda$. The $r$-independent
coefficients $c_{\mu,\nu}$ are a {\it single} realization of complex, random,
and uncorrelated numbers: their real and imaginary part are both chosen
corresponding to a Gaussian distribution with the mean $0$ and the variance
$1/2$. Of particular interest is the probability for finding the particle
somewhere within the $r$th local module. The associated local current is
$j_r = \imath \, \lambda \, m_r + \text{H.c.}$ with a form very similar to
$v_r$, e.g., almost completely random (apart from the translation invariance
and the necessary Hermitian property).

The modular quantum system is one of the few quantum models which have been
reliably shown to exhibit diffusion with a finite diffusion constant, reading
$D_\mathrm{th}^{\mathrm{w}}(t>\pi/\delta\epsilon)=2\pi\lambda^2n/\delta\epsilon$
for weak coupling \cite{steinigeweg2007} and
$D_\mathrm{th}^{\mathrm{s}}=(t>1/\lambda\sqrt{2n})=\lambda\sqrt{\pi n/2}$
for strong coupling \cite{steinigeweg2011-1}. One might expect a
finite Burnett coefficient due to both the presence of diffusion and the random
elements of the current. For instance, because $J^2$ is close to an identity
matrix, one may be tempted to factorize as $\langle J(0) J(t_1) J(t_2) J(t_3)
\rangle = f(t_3) f(t_1-t_2)$, then allowing for a finite Burnett coefficient.
But the latter factorization already fails when all time arguments are equal. In
fact, $\langle J^4 \rangle = 2 f(0)^2 = 8 \lambda^4 n^2$, resulting from an
additional {\it coherent} sum over the off-diagonal elements of $J^2$. Instead,
fulfilling the static property, we may choose the {\it unbiased} factorization
of $\langle J(0) J(t_1) J(t_2) J(t_3) \rangle$ into $2/3[ f(t_1) f(t_2-t_3)
+ f(t_2) f(t_1-t_3) + f(t_3) f(t_1-t_2)]$, yielding the relation $I_4(t) = 2/3
I_{2-2}(t)$ between the two contributions to $B(t)$. Therefore, noting that
$\chi = 1$, this choice leads to the {\it linearly} increasing Burnett coefficient
\begin{equation}
B(t > \tau) \simeq -\frac{2}{3} D_\mathrm{th}^2 t \, .
\end{equation}
For verification, we present in Fig.~\ref{FigIV} numerical results on $D(t)$ and
$B(t)$.
 Because the linear growth of the Hilbert space with $L$ is compensated by
translation invariance, the dimension of a momentum $k$-subspace is only set by the level
number $n$, chosen as $n=500$ to ensure a sufficient number of states. Since we
do not find a significant dependence on $k$, Fig.~\ref{FigIV} depicts numerical
results for a single $k$-subspace. The quantitative agreement with the theoretical
predictions on $D(t)$ and $B(t)$ clearly demonstrates a linearly increasing Burnett
coefficient despite the existence of a diffusion constant, which is in agreement
with the previous results on spin chains, but much clearer due to computational
simplicity of the model.

{\it Conclusion.}---
In this Letter we presented extensive numerical and theoretical investigations of
Burnett coefficients in quantum chains. We conjectured and supported the general
statement that in the thermodynamic limit Burnett coefficients diverge linearly,
$B(t) \propto t$, in diffusive regimes with finite diffusion constants. Recall that
this linear divergence is still consistent with diffusion
but causes Burnett coefficients to be a relevant correction at arbitrary small
momentum. In ballistic regimes with positive Drude weights, on the other hand, we
demonstrated the cubic divergence $B(t) \propto t^3$. This behavior is 
remarkably different than for Lorentz billiard type classical systems and and calls for
a deeper theoretical analysis.

\acknowledgments

Financial support by the grant P1-0044 of Slovenian research agency is gratefully acknowledged.


\begin{thebibliography}{99}
\bibitem{susceptibility} We define the ``static susceptibility'' as the square of the
conserved quantity \cite{steinigeweg2011-1},
e.g.~$\chi = \langle (S^z)^2 \rangle$ for spin transport with $S^z$ being the
conserved magnetization.
\bibitem{algorithm}
The time order of $I_4(t)$ can be determined by numerically solving a
set of coupled differential equations for auxiliary operators $M_i(t)$,
$$
I_4(t) = \frac{2}{\chi} \langle J(0) M_3(t) \rangle \, , \quad \dot{M}_i(t) =
J(t) M_{i-1}(t) \, ,
$$
where $M_0(t) = 1$. Here, we use full exact diagonalization of $H$ to evaluate $J(t)$
in the Heisenberg picture and a 4th order Runge-Kutta integration.
\end{thebibliography}
\end{document}